\documentclass[reprint,superscriptaddress,
 amsmath,amssymb,
 aps,
 prl,
]{revtex4-2}

\usepackage{graphicx} 
\usepackage{dcolumn}
\usepackage{xcolor} 
\usepackage{bm}
\usepackage[colorlinks=true, linkcolor=blue, citecolor=blue, urlcolor=blue]{hyperref}
\usepackage[english]{babel}

\makeindex

\begin{document}

\preprint{APS/123-QED}

\title{Constraining Nuclear Mass Models Using r-process Observables with Multi-objective Optimization}

\author{Mengke Li}
\email{mengkel@berkeley.edu}
\affiliation{Department of Physics, University of California, Berkeley, Berkeley, California, 94720, USA}
\affiliation{Department of Physics and Astronomy, University of Notre Dame, Notre Dame, Indiana 46656, USA}

\author{Matthew Mumpower}
\affiliation{Theoretical Division, Los Alamos National Laboratory, Los Alamos, NM, 87545, USA}
\affiliation{Center for Theoretical Astrophysics, Los Alamos National Laboratory, Los Alamos, NM, 87545, USA}

\author{Nicole Vassh}
\affiliation{TRIUMF, 4004 Wesbrook Mall, Vancouver, British Columbia V6T 2A3, Canada}

\author{William Samuel Porter}
\affiliation{Department of Physics and Astronomy, University of Notre Dame, Notre Dame, Indiana 46656, USA}

\author{Rebecca Surman}
\affiliation{Department of Physics and Astronomy, University of Notre Dame, Notre Dame, Indiana 46656, USA}

\date{\today}

\begin{abstract}


Predicting nuclear masses is a longstanding challenge. One path forward is machine learning (ML) which trains on experimental data, but can suffer large errors when extrapolating toward neutron-rich species. In nature, such masses shape observables for the rapid neutron capture process (r-process), which in principle could inform ML models. Here we introduce a multi-objective optimization approach using the Pareto Front algorithm. We show that this technique, capable of identifying models which generate r-process abundances aligning with both Solar and stellar data, is a promising method to select ML models with reliable extrapolation power.

\end{abstract}


\maketitle

\section{\label{sec:intro}Introduction}

The nuclear mass is the fundamental ground state property of a nuclear species. Predicting exactly how masses evolve for species away from the valley of stability remains one of the key unanswered questions of modern nuclear physics \citep{NSAC2023}. Experimentally, there has been impressive progress in approaching exotic nuclear species in rare isotope facilities \citep{Madurga_2012, Lorusso_2015, Atanasov_2015,Vilen_2018}. In cases where experimental data is lacking, nuclear physics models are employed to predict the properties of unmeasured nuclei \citep{Lunney+2003}.

Several theoretical approaches, including phenomenological and macroscopic-microscopic models, are used to predict nuclear masses \citep{DZ_1995, Koura_2005, Moller_2012, HFB-32}. While these models generally agree well in regions where experimental data is available, they diverge significantly in extrapolated regions. With advancements in computing power, many machine learning (ML) algorithms have been increasingly applied to model and extrapolate nuclear masses far from stability \citep{Lovell_2020, Gao-ML, Niu_2022, mumpower2023bayesian, Yuksel_2024}. The advantage of ML models lies in their data-driven nature, enabling global optimization of residuals for highly precise predictions. Furthermore, ML models can quantify prediction uncertainties, offering valuable insights into the reliability of predictions in unexplored regions. Additionally, training a new ML-based model is significantly faster than developing a classical model (which is subject to both limitations in computational expense and physical assumptions), enabling the rapid generation of data using state-of-the-art techniques and architectures.

In order to evaluate a nuclear mass model, one approach is to compare its predictions with available experimental measurements. 
Among the approximately 9000 nuclei theorized to exist, experimental masses are available for only about 2500 \cite{AME2020}. 
As a result, generating a model that describes measured data well does not imply well-behaved extrapolations to unmeasured regions, particularly for neutron-rich nuclei far from stability. 

While the vast majority of exotic neutron-rich nuclear species have not been produced in the laboratory, they are accessed in nature through rapid neutron capture ($r$) process nucleosynthesis. In the $r$ process, a sequence of rapid neutron captures and beta decays builds increasingly heavy nuclei along a nucleosynthesis avenue that runs roughly parallel to the valley of stability on the neutron-rich side. The temperature and density conditions are expected to be sufficiently extreme to result in an equilibrium between neutron captures and photodissociations---($n$,$\gamma$)-($\gamma$,$n$) equilibrium---where the abundances along an isotopic chain are set by a Saha equation that depends exponentially on the nuclear mass differences. Therefore,
observed r-process abundances, shaped by the properties of nuclei beyond experimental reach in extreme astrophysical environments, can provide a critical constraint in this context because they encode rich information about the underlying nuclear physics. 
Using predicted nuclear masses in r-process simulations, therefore, serves as a powerful method to test and quantify the quality of these extrapolations \citep{Li_2024}. This approach connects theoretical mass models to astrophysical observables, offering insight into the behavior of nuclei in the most neutron-rich conditions.

To identify nuclear mass models that not only accurately reproduce experimental data but also offer reliable extrapolations, this work presents the first application of a multi-objective optimization algorithm to select suitable ML-based nuclear mass models. In this letter, we first introduce the Pareto Front (PF) optimization algorithm, followed by a discussion of the training process for the ML mass models and their impact on r-process simulations. Then, we explore the application of the PF algorithm in selecting mass models. Finally, we compare the properties of the selected models with those from the larger set of models and provide our concluding remarks.

\section{\label{sec:method}Method}
The PF algorithm is a method for addressing multi-objective optimization problems, where multiple, potentially conflicting, conditions must be optimized simultaneously. Instead of reducing the solution to a single best option, the PF algorithm identifies a set of Pareto-optimal solutions, each representing a trade-off between the objectives. Solutions on the Pareto Front are considered optimal because no other solution can improve one objective without compromising another \citep{PF_2005}. Formally, a decision vector $\vec{\boldsymbol{u}}$ is said to Pareto-dominate another vector $\vec{\boldsymbol{v}}$ if and only if:\\
1. $\vec{\boldsymbol{u}}$ is at least as good as $\vec{\boldsymbol{v}}$ in all objectives: 
\begin{equation}
    \forall i \in \{1, \dots, N\}, f_i(\vec{\boldsymbol{u}}) \leq f_i(\vec{\boldsymbol{v}})
\end{equation}
2. $\vec{\boldsymbol{u}}$ is strictly better than $\vec{\boldsymbol{v}}$ in at least one objective:
\begin{equation} 
\exists j \in \{1, \dots, N\} : f_j(\vec{\boldsymbol{u}}) < f_j(\vec{\boldsymbol{v}})
\end{equation}

To apply this method to ML-based mass model selection, each vector represents the predicted masses across the entire nuclear chart. In this work, we define three objective functions. The first ($f_1$) evaluates the model’s accuracy in predicting experimental nuclear masses. The second ($f_2$) assesses its ability to reproduce solar r-process isobaric abundances, while the third ($f_3$) measures its ability to replicate observed stellar elemental r-process abundances. 

When training ML nuclear mass models, similar to our previous work \citep{mumpower2023bayesian}, we utilize the Mixture Density Network (MDN) framework \cite{MDN} to predict nuclear masses, with input data being a hybrid of experimental and theoretical values in regions devoid of measurements. Recognizing that training outcomes are influenced by factors such as the choice of training samples, input feature space, and neural network architecture, we generate a diverse pool of ML mass models by varying these parameters while ensuring uniform training durations for all models. 

Building on insights from earlier studies, which recognized the critical role of mass differences when linking masses and r-process abundances \citep{Mumpower_2017, Vassh_2021}, we enhance the training process by incorporating mass differences, such as neutron separation energies, as supplementary constraints to train the mass models. This modification is designed to improve the models’ extrapolation ability to reproduce physical trends, and overall predictive performance.

When each ML mass model is well trained, its predictions are used to simulate r-process nucleosynthesis. Specifically, we calculate the neutron separation energy ($S_n$) for all nuclei using the new mass data, which directly influences photodissociation rates via detailed balance, playing a critical role in the r-process \citep{Kajino_2019, Li_2022}. The calculated $S_n$ are then input into the nuclear network code PRISM \citep{Mumpower_2018} to perform the calculation. For all reaction rates, we employ a combination of experimental data (when available) and theory calculations \citep{Mumpower_2016bip, MOLLER20191, Kawano_2016, Vassh_2020} to ensure comprehensive data coverage of the nuclear chart.
After obtaining each mass model and applying it to the r-process simulations, we calculate the root-mean-square ($\rm{RMS}$) error and $\chi^2$ error of the mass model relative to AME2020 dataset \citep{Wang_2021} to assess its predictive power by using the following equations:
\begin{equation}
    \sigma_{\rm{RMS}} = \sqrt{\frac{1}{N}\sum_{i}(t_i - d_i)^2}
\end{equation}
and 
\begin{equation}\label{eq:chi2}
    \chi^2 = \sum_i \frac{(t_i - d_i)^2}{\sigma_i^2}
\end{equation}

Here, $t_i$ represents the atomic mass predicted by the ML model, $d_i$ denotes the atomic mass from the AME2020 dataset, and $\sigma_i$ corresponds to the uncertainty associated with each experimental atomic mass. The calculated $\chi^{2}$ for masses serves as the first objective function, $f_1$. Next, we compute the $\chi^{2}$ errors using Eq. (\ref{eq:chi2}) for the simulated isobaric abundances $Y(A)$, with $A$ larger than 120, relative to solar data \citep{Beer_1997} and elemental abundances $Y(Z)$, with $Z$ ranging from 56 to 79 to cover the main r-process elements, relative to a metal-poor r-process-enhanced halo star HD 222925 that has a nearly complete r-process chemical inventory \citep{Roederer_2022}. These two quantities serve as the second and third objective functions, $f_2$ and $f_3$, respectively. Here, $t_i$ represents the simulated isobaric or elemental abundances obtained using each ML mass model, $d_i$ denotes the corresponding solar or stellar abundances, and $\sigma_i$ is the observed uncertainty in the abundances. Thus, each ML mass model is characterized by three unique properties: $\chi^2$ values for masses, $Y(A)$, and $Y(Z)$. During the optimization process, we evaluate the performance of each model across these three objectives using our PF algorithm. 

\section{\label{sec:PF_results}Results}

\subsection{\label{sec:mass_models_r}Comparing the properties of different ML mass models} 

As a first step, we generate ML mass models for use in our PF analysis. The performance of these models in reproducing AME masses and solar $r$-process isotopic abundances are shown in Fig.~\ref{fig:mass_r}.
The upper panel displays the RMS error relative to the AME2020 database. As described in the method section, there are two different categories of ML mass models: those trained with and without mass differences as additional constraints. The grey distribution represents MDN models that were trained without incorporating mass differences as constraints. In total, 600 mass models were trained, each differing in hyper-parameters as described in our previous work \citep{Li_2024}. The RMS error ranges from 0.18 to 0.5 MeV, which is comparable to other mass models in the literature, indicating the success of these models in predicting experimentally available masses.

The RMS errors for MDN models trained with mass differences are shown in green. Here, 
in total 200 models were trained, each differing in hyper-parameters and the specific mass differences used (e.g., one-neutron separation energies $S_n$, $\beta^-$-decay energies $Q_{\beta}$, $\alpha$-decay energies $Q_{\alpha}$, among others). This set of 200 gives RMS mass deviations ranging from 0.25 to 0.48. It is evident that the distribution range of ML models without mass differences in the training exhibits a wider spread compared to those trained with mass differences. This indicates greater uncertainty in models trained without mass differences as additional constraints. While some models trained without mass differences achieve lower RMS values, suggesting a good match to the available data near stability, these cases do not {necessarily preserve expected mass difference trends away from stability. This
underscores the need for an evaluation of the model’s performance beyond matching existing data.

\begin{figure}[h]
    \centering
    \includegraphics[width=1\linewidth]{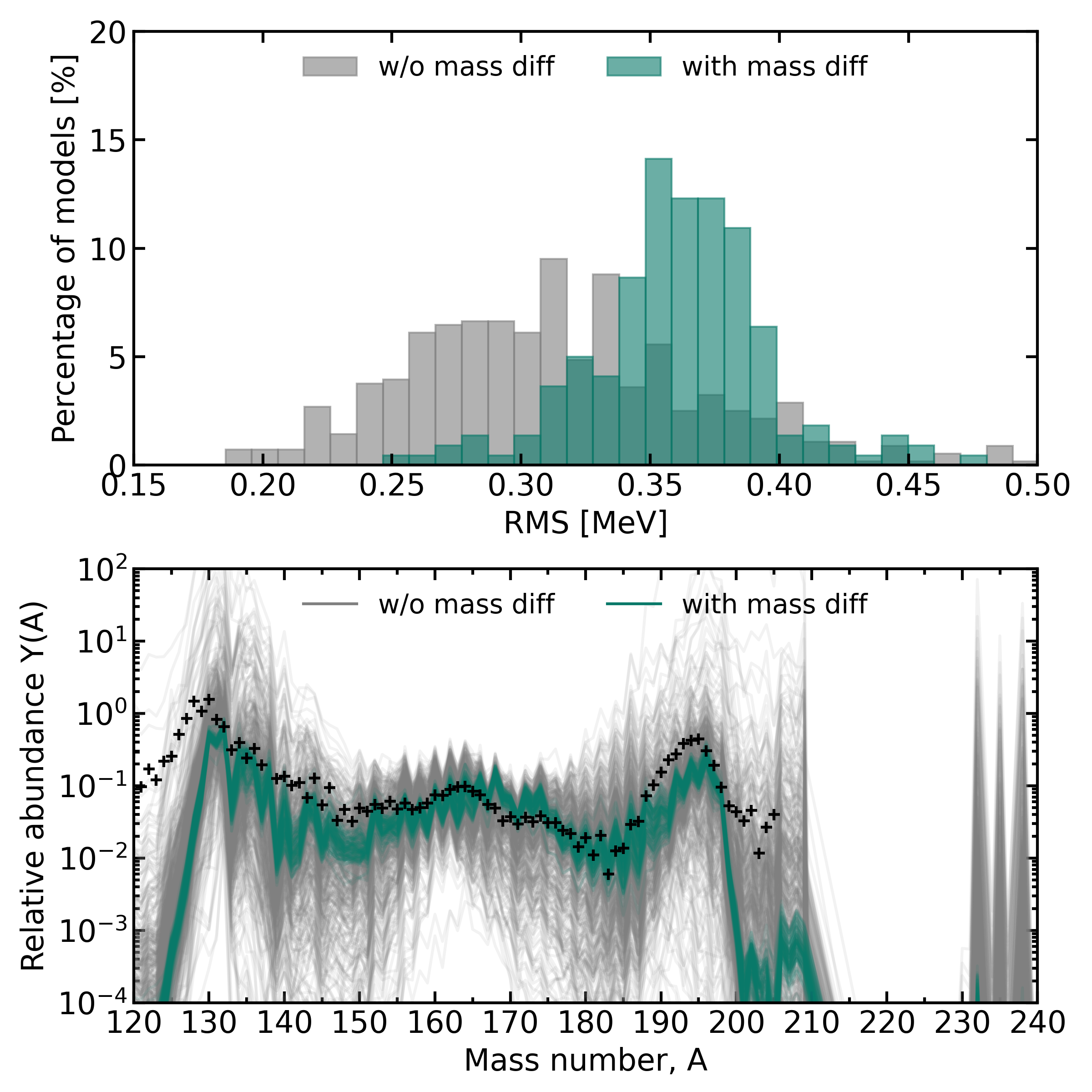}
    \caption{The upper panel shows the distribution of the RMS for all well-trained mass models, with green and grey squares representing MDN models trained with and without mass differences, respectively. The lower panel displays the corresponding simulated r-process abundance patterns under a hot wind condition, compared to solar residuals from \cite{Beer_1997}.}
    \label{fig:mass_r}
\end{figure}

To further investigate which models provide more reliable extrapolations for neutron-rich nuclei, we apply these models to r-process simulations, which can serve to inform mass predictions in the absence of experimental data. The resulting abundance patterns calculated with hot wind conditions {\citep{Zhu_2018}} are shown in the lower panel of Fig.~\ref{fig:mass_r}. Green and grey lines represent the abundance patterns from MDN models trained with and without mass differences, respectively. The narrower green band, with distinct and consistent r-process characteristic features, compared to the wider grey band, highlights the constraining power that introducing mass differences in training can provide, particularly in informing extrapolations outside of measured data.

\subsection{\label{sec:PF_mass_models}PF algorithm in constraining ML mass models}

The previous section demonstrates the advantage of incorporating mass differences when training the MDN models. Here we describe how $r$-process simulations are used to provide additional constraints on our ML models, 
to ensure that the extrapolated masses are physical. Among the 200 well-trained models, to select the most robust and predictive ones, we apply them to a broader set of $r$-process calculations.
We employ a set of 10 representative trajectories from a simulation of neutron-star merger disk ejecta \citep{just_2015}, covering a range of neutron-rich conditions.
The normalized r-process abundances $Y(A)$ and $Y(Z)$ are calculated for each simulation using different mass models. By comparing $Y(A)$ to solar data \citep{Beer_1997} and Y(Z) to stellar data \citep{Roederer_2022}, we obtain the $\chi^2$ $Y(A)$ and $\chi^2$ $Y(Z)$ values for each simulation. As a result, each MDN model has three distinct properties: $\chi^2$ masses, $\chi^2$ $Y(A)$, and $\chi^2$ $Y(Z)$. 

To choose those with low $\chi^2$ mass values and producing the r-process patterns most in line with Solar and stellar data, we employ the PF algorithm for multi-objective optimization. Fig. \ref{fig:PF_points} presents the $\chi^2$ values for each MDN model, shown as green squares. The PF algorithm selects models with low values in any of the three objectives, represented by pink squares. Since we are using three objectives to describe each MDN mass model, the Pareto front set forms a curved 2D surface within a 3D space, which can be found at our OSF page \citep{pf_3d}. Fig.~\ref{fig:PF_points} shows 2D projections of this 3D space: the left panel plots $\chi^2$ masses against $\chi^2$ $Y(A)$, and the right panel plots $\chi^2$ $Y(Z)$ against $\chi^2$ $Y(A)$. Both projections clearly indicate that the PF set occupies the lower left corner of the plots, demonstrating that the selected models have lower $\chi^2$ values for masses, $Y(A)$, and $Y(Z)$. These selected models balance the three metrics so that there is no one metric considered to be optimal among others; together, they outperform the remaining MDN models.

\begin{figure}[h!]
    \centering
    \includegraphics[width=1\linewidth]{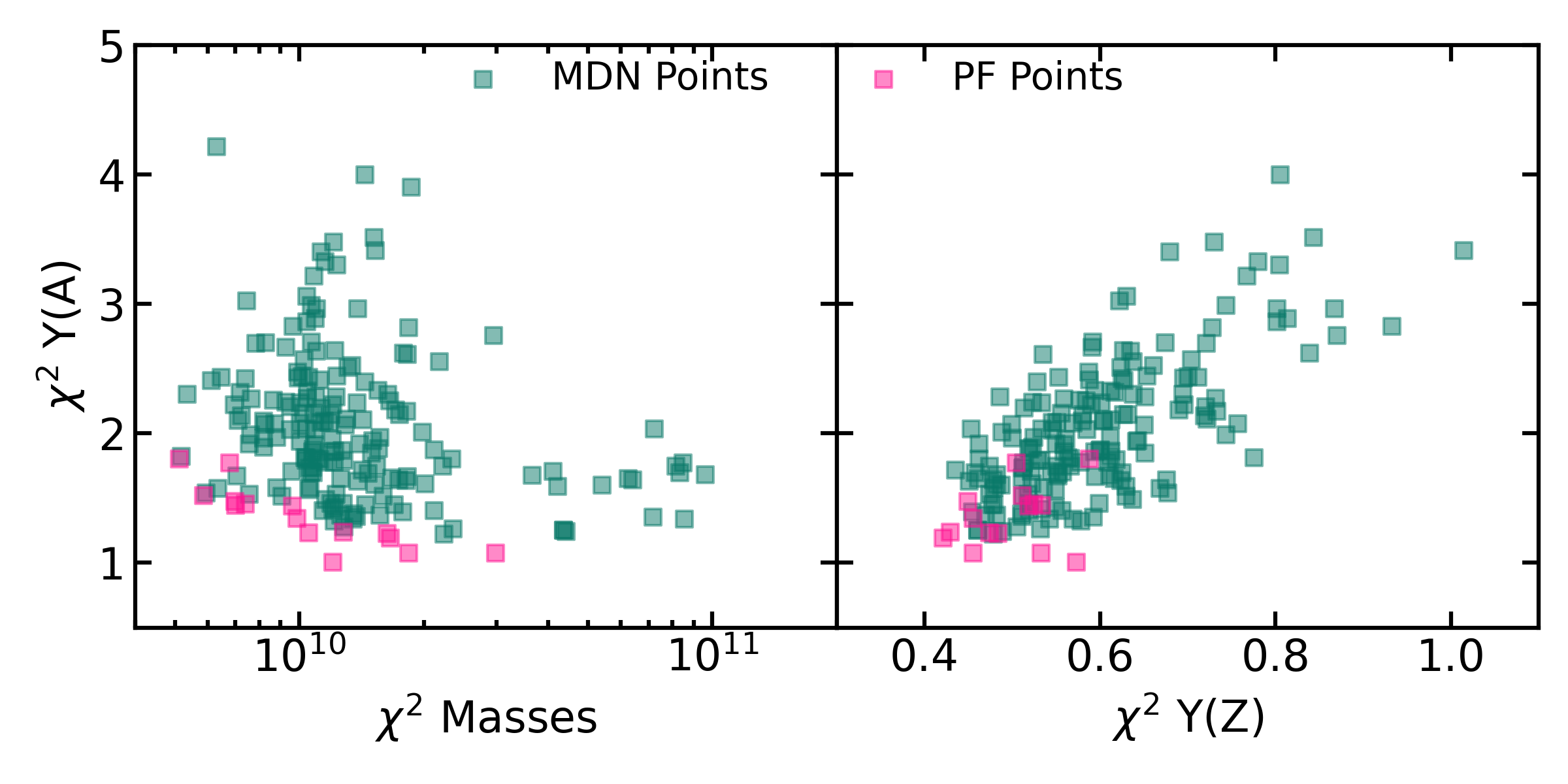}
    \caption{The left panel shows the models' properties of $\chi^2$ masses vs. $\chi^2$ $Y(A)$ with the green and pink squared standing from the MDN model with mass difference in training and the PF set. The right panel show the models' properties of $\chi^2$ $Y(Z)$ vs. $\chi^2$ $Y(A)$.}
    \label{fig:PF_points}
\end{figure}

Fig.~\ref{fig:PF_ya_yz} presents a distribution of abundance patterns with $Y(A)$ shown on the top panel and $Y(Z)$ on the bottom panel. It is evident from the upper panel that the general isobaric abundance patterns are aligned well with the solar pattern, with the characteristic peaks positioned appropriately, indicating reasonable atomic mass extrapolations for neutron-rich nuclei. It is worth noting that the variations in the abundance distributions
are not uniform across different mass numbers. Except for some outliers in the second ($A \approx$ 130) and third ($A \approx$ 195) peak region, the variations in the deformed nuclei and after the third peak region are greater than other regions, indicating the larger variation in atomic mass extrapolations related to the formation of these nuclei. A similar phenomenon is observed in the lower panel, which shows the distribution of the elemental abundances. 

To compare the abundance patterns between the PF-selected models and the MDN models, the upper panel of Fig.~\ref{fig:PF_ya_yz} illustrates the isobaric abundance patterns for both model sets, with the MDN models depicted by green lines and the PF models by pink lines. 

The wider spread of the abundances in MDN models indicates larger variability in the mass predictions, particularly in the extrapolated region. In contrast, the narrower spread of the abundances in PF models underscores the algorithm’s effectiveness in selecting models with improved extrapolation capabilities for neutron-rich nuclei. This suggests that the PF algorithm effectively helps the model selection process, yielding more physically plausible results. A similar feature is observed in the elemental abundance distributions shown in the lower panel, where the PF models exhibit tighter distributions, further demonstrating the algorithm’s capacity to produce more reasonable abundance patterns.

\begin{figure}[h!]
    \centering
    \includegraphics[width=1\linewidth]{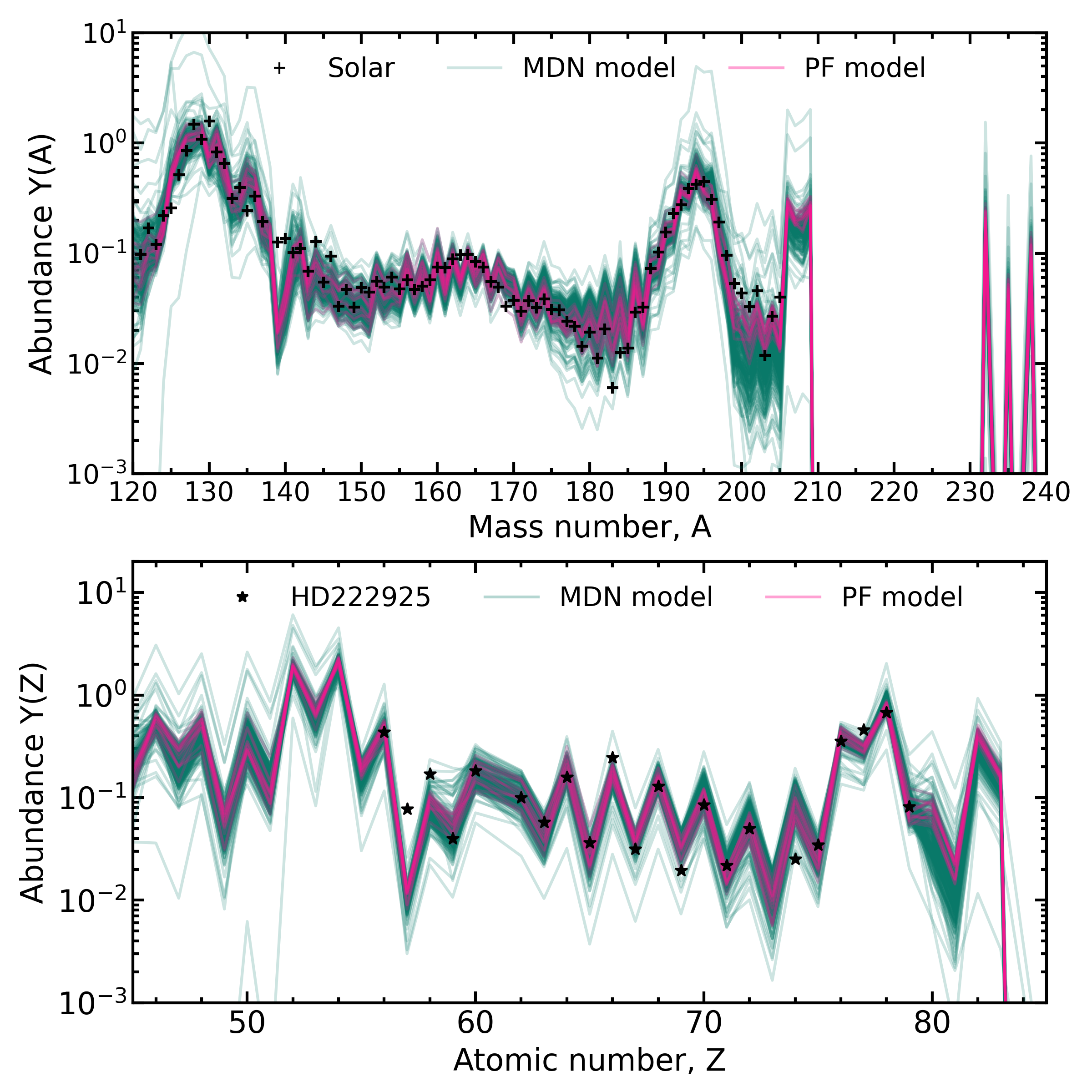}
    \caption{The upper panel shows the distribution of the simulated isobaric abundances for all the MDN models (green lines) and the Pareto front set (pink lines), while the lower panel presents elemental abundances for all different models. The black crosses are the solar r-process residuals \citep{Beer_1997}, while the black triangles are data from a metal-poor r-process-enhanced halo star HD222925 \citep{Roederer_2022}.}
    \label{fig:PF_ya_yz}
\end{figure}

\subsection{\label{sec:PF_Sn}PF algorithm in constraining neutron separation energies}

Another quantity that can be used to gauge the quality of extrapolations is one-neutron separation energy ($S_n$). Not only is it an important quantity relevant to closed-shell signatures, pairing effects, and the boundaries of the nuclear landscape, it also sets the equilibrium r-process path through the calculation of the photo-dissociation rate via detailed balance. Accurate extrapolation of separation energies is vital for both nuclear physics and astrophysical applications. Key features of neutron separation energies include a decreasing trend along an isotopic chain as isotopes become more neutron-rich and an odd-even staggering effect observed between isotopes with odd and even neutron numbers.

To assess the quality of the $S_n $ extrapolation between the general MDN models and those selected by the PF algorithm, Fig.~\ref{fig:PF_sn} presents the $S_n$ values for the Tin isotopic chain as an example. The distribution of $S_n$ values from the MDN models without and with mass difference as an additional constraint are shown in grey and green bands, respectively. In the experimentally-known region, these distributions nearly overlap with minimal variation, which is expected given their low root-mean-square (RMS) deviations relative to AME2020. However, as the neutron number increases toward the drip line, incorporating mass differences in training significantly constrains the extrapolation, narrowing down the deviations.

The PF-selected models, represented by pink lines, show an additional improvement in the extrapolated values. These models exhibit a clear odd-even staggering effect up to and beyond the neutron drip line. In contrast, many of the models constrained only by mass differences, shown in the green band, lose this property. Notably, similar constraints are also effective on the proton-rich side when the neutron number is less than $50$, where we observe a progressive narrowing from the broad gray band to green, and further to pink.
This demonstrates that the PF algorithm effectively selects ML mass models that preserve physically meaningful extrapolations, maintaining both the expected decreasing trend and odd-even effects. This behavior is consistently observed across all isotopic chains, with this case serving as a representative example.
 
\begin{figure}[h!]
    \centering
    \includegraphics[width=1\linewidth]{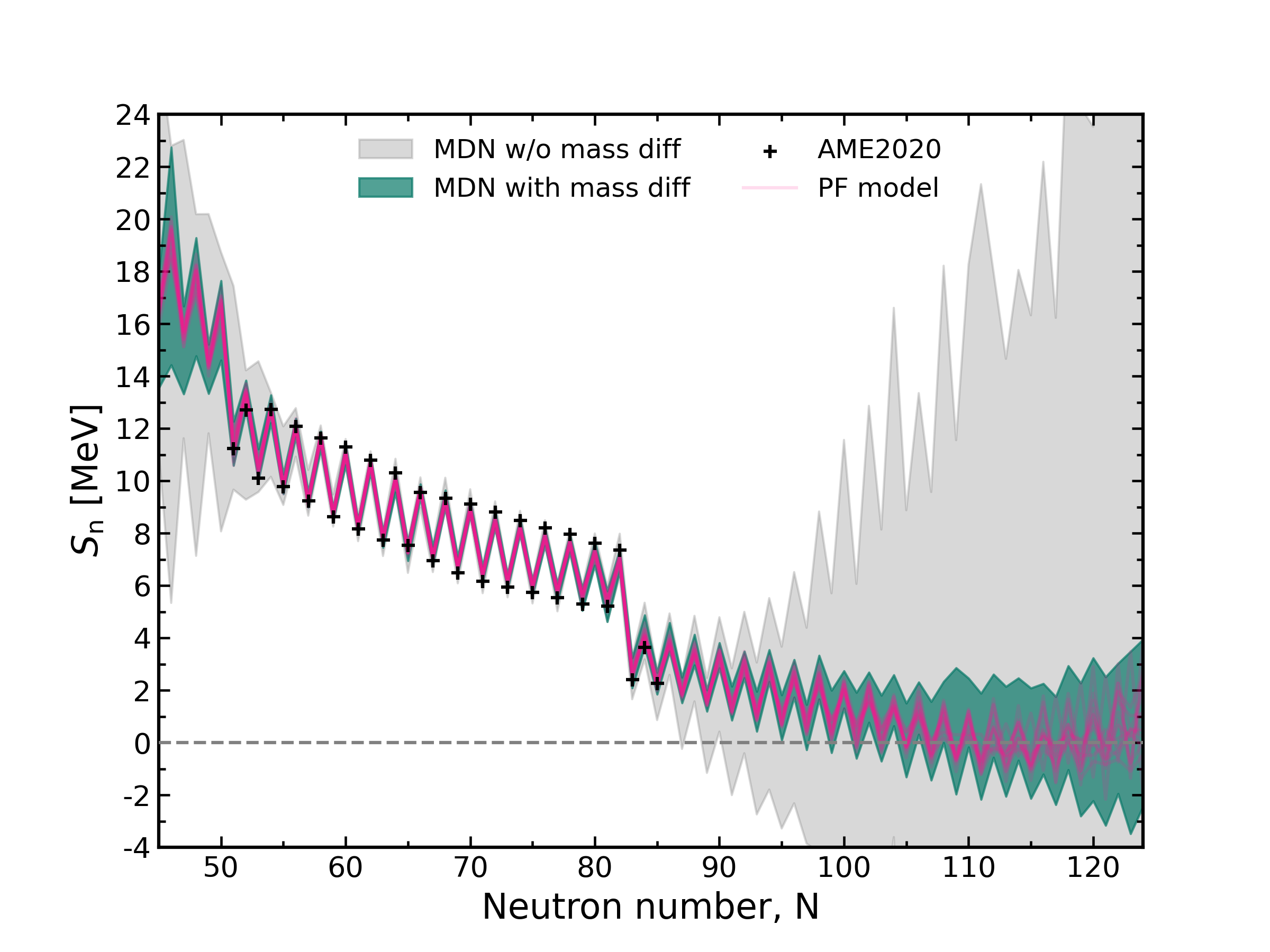}
    \caption{The grey and green bands show the distribution of one neutron separation energies ($S_n$) in the Tin ($Z = 50$) isotopic chain calculated with MDN mass models without and with mass difference constraints, respectively, while the pink lines highlight the calculated $S_n$ in the PF set. The dashed horizontal grey line shows the neutron drip line ($S_n=0$). }
    \label{fig:PF_sn}
\end{figure}

\section{\label{sec:conclusion}Conclusion and Discussion}

In conclusion, we have demonstrated that incorporating mass differences as additional constraints in the training of ML nuclear mass models maintains their predictive accuracy and enhances their ability to make physically meaningful extrapolations across the nuclear chart. This improvement is evident in the simulated r-process abundance patterns. Utilizing MDN models trained with mass difference and the corresponding simulated r-process abundance patterns, we applied the PF algorithm to select mass models that exhibit lower $\chi^2$ values across three key metrics: nuclear masses, isobaric abundances $Y(A)$, and elemental abundances $Y(Z)$. The models chosen by the PF algorithm produce narrower distributions of RMS values and r-process abundance patterns, reflecting more refined predictions, especially in challenging regions of the nuclear landscape, such as deformed nuclei and those near the neutron drip line. Furthermore, the improvement in the predictive power of extrapolations is also demonstrated in the accurate reproduction of key nuclear properties such as the one-neutron separation energies. 

This study marks the first application of the PF algorithm to constrain ML nuclear mass models using both experimental mass data and r-process observables, highlighting its potential as a robust tool for model selection. The multi-objective optimization framework provided by the PF algorithm is versatile and can be extended to any physical models characterized by multiple competing properties. This could include applications in other areas of nuclear physics, astrophysics, and beyond, where selecting models across multiple criteria is necessary.

\begin{acknowledgments}
M.~L. and R.~S. acknowledge support from the
Network for Neutrinos, Nuclear Astrophysics and Symmetries (N3AS), through the National Science Foundation Physics Frontier Center award No. PHY-2020275.
M.~R.~M. acknowledges support from the Laboratory Directed Research and Development (LDRD) program of Los Alamos National Laboratory (LANL) project number 20240004DR. 
This research was supported in part by LANL through its Center for Space and Earth Science (CSES). 
CSES is funded by LANL’s LDRD program under
project number 20210528CR. 
LANL is operated by Triad National Security, LLC, for the National Nuclear Security Administration of U.S. Department of Energy (Contract No. 89233218CNA000001). 
N.~V. acknowledges the support of the Natural Sciences and Engineering Research Council of Canada (NSERC).
W.~S.~P. acknowledges support from the National Science Foundation under Grant No. PHY-2310059. R.~S. additionally acknowledges support from the U.S. Department of Energy under Grant Nos.~DE-FG02-95-ER40934, LA22-ML-DE-FOA-2440, and DE-SC00268442 (ENAF).
\end{acknowledgments}

\bibliography{PF}

\end{document}